\documentclass[a4paper,12pt]{article}
\usepackage{amsmath}
\usepackage{amsfonts}
\usepackage{amssymb}
\usepackage{latexsym}
\usepackage{epsfig}
\usepackage{graphicx}
\usepackage{oldgerm}
\usepackage{theorem}

\def\r{\mib{r}}
\def\k{\mib{k}}
\def\q{\mib{q}}
\def\vsigma{\mib{\sigma}}
\def\v{\mib{v}}

\def\B{\mib{B}}
\def\K{\mib{K}}
\def\A{\mib{A}}
\def\J{\mib{J}}
\def\E{\mib{E}}

\def\N{\mathbb{N}}

\def\R{\mathbb{R}}

\theorembodyfont{\itshape}

\newcommand{\mib}[1]{\mbox{\boldmath $#1$}}
\newcommand{\SSC}[1]{\section{#1}\setcounter{equation}{0}}

\begin{document}

\title{\bf 
Height correlation of rippled graphene \\
and Lundeberg-Folk formula for magnetoresistance
}
\author{
Kazuyuki Gemma and 
Makoto Katori
\footnote{
Department of Physics,
Faculty of Science and Engineering,
Chuo University, 
Kasuga, Bunkyo-ku, Tokyo 112-8551, Japan;
e-mail: katori@phys.chuo-u.ac.jp
}}
%%%%%%%%%%%%%%%%%%%%%%%%%%%%%%%%
\date{2 December 2013}
%%%%%%%%%%%%%%%%%%%%%%%%%%%%%%%
\pagestyle{plain}
\maketitle
\begin{abstract}
Application of an in-plane magnetic field to
rippled graphene will make the system be
a plane with randomly distributed
vector potentials.
Massless Dirac fermions carrying charges on graphene
are scattered by the vector potentials
and magnetoresistance is induced
proportional to the square of amplitude
of in-plane magnetic field $B_{\parallel}^2$.
Recently, Lundeberg and Folk proposed a formula showing
dependence of the magnetoresistance on
carrier density, in which
the coefficient of $B_{\parallel}^2$
is given by a functional of the height-correlation
function $c(r)$ of ripples.
In the present paper, we give exact and explicit
expressions of the coefficient
for the two cases such that $c(r)$ is (i) exponential
and (ii) Gaussian.
The results are given using well-known special functions.
Numerical fitting of our solutions
to experimental data were performed.
It is shown that 
the experimental data are well-described
by the formula for the Gaussian height-correlation
of ripples in the whole region of carrier density.
The standard deviation $Z$ of ripple height
and the correlation length $R$ of ripples 
are evaluated,
which can be compared with direct experimental measurements.
\end{abstract}

%%%%%%%%%%%%%%%%%%%%%%%%%%%%%%%%%%%%%%%%%%%%%%%%%%%%%%%%%%
%%%  SEC1 %%%%%%%%%%%%%%%%%%%%%%%%%%%%%%%%%%%%%%%%%%%
%%%%%%%%%%%%%%%%%%%%%%%%%%%%%%%%%%%%%%%%%%%%%%%%%%%%%%%%%%
\SSC{Introduction and main results}
\label{chap:intro} 
%%%%%%%%%%%%%%%%%%%%%%%%%%%%%%%%%%%%%%%%%%%%%%%%%%%%%%%%%%

Graphene, a one-atom-thick allotrope of carbon
\cite{NGMJZDGF04,ZTSK05}, 
has attracted much attention of theoretical physicists,
since its electronic excitations are described
by two-dimensional (2D) {\it massless Dirac fermion}
(see, for instance, \cite{CGPNG09,DSAHR11}).
Let $v_{\rm F}$ be the Fermi velocity.
Then, for a free fermion, the 2D massless Dirac equation
is given by
\begin{equation}
-i \hbar v_{\rm F} \left(
\begin{array}{cc}
0 & \partial_x-i \partial_y \cr
\partial_x+i \partial_y & 0
\end{array} \right)
\psi(\r)=E \psi(\r),
\label{eqn:Dirac1}
\end{equation}
where $i=\sqrt{-1}$,
$\partial_x=\partial/\partial x$,
$\partial_y=\partial/\partial y$,
and $\psi(\r)$ is a two-component wave function
with the particle position $\r=(x,y)$.
For a wave-number vector $\k=(k_x, k_y)$,
the eigenvalues of energy are given by
$E=E_{\pm} = \pm \hbar v_{\rm F} k$, $k=|\k|$,
where the $\pm$ signs correspond to 
two values of the helicity of fermion.
The wave functions are determined as \cite{CGPNG09}
\begin{equation}
\psi_{\pm}(\r; \k)
=\langle \r | \k \rangle
=\frac{1}{\sqrt{2V}} e^{i \k \cdot \r}
\left( \begin{array}{cc}
e^{-i \theta_k/2} \cr
\pm e^{i \theta_k/2}
\end{array} \right),
\label{eqn:Dirac2}
\end{equation}
where $V$ is the volume of the system
and 
\begin{equation}
\theta_k=\arctan \left(\frac{k_y}{k_x}\right).
\label{eqn:Dirac3}
\end{equation}

Although the basic description of the system is very simple
as shown above, physics of graphene is extremely rich,
since the 2D Dirac equation is much modified
by external electric and magnetic fields 
applied to the graphene plane as well as by
changing geometric and topological features
of the 2D surface \cite{CGPNG09,DSAHR11}.
As a matter of fact, an ultrathin membrane is
unstable in the 3D real space and 
in suspended graphene and in graphene placed
on a substrate nanometer-scale
corrugations, which are simply called {\it ripples}
\cite{FLK07},
are observed 
\cite{MGKBNR07,ICCFW07,Stolyarova07,Stoberl08,Tikhonenko08,Lui09,Geringer09}.
Since spatial distribution of such microscopic
structures on the surface cannot be controlled,
even when we apply uniform electric and/or
magnetic fields, the rippled graphene will behave
as a 2D system with randomly distributed
scalar and vector potentials on it.
Effect of ripples on electronic transport
on graphene is an interesting subject
both from theory and experiment 
\cite{CGPNG09,DSAHR11,Morozov06,KG08,Morozov08,Deshpande09}.

Recently, magnetotransport measurements
on rippled graphene have been reported,
where a magnetic field is applied parallel to the 
averaged 2D plane of the graphene \cite{LF10,LF10b,WS11,Shi12}
(see also \cite{WS12} for bilayer graphene).
If the surface is a perfect plane, such 
an in-plane magnetic field $\B_{\parallel}$ 
does not affect any motion of 2D electron,
since electronic motion couples only to
the component of magnetic field normal to
the 2D plane.
But on real graphene, the magnetic field 
becomes to include a component normal to the surface
depending on the local slope around ripples.
Then the system has inhomogeneous out-of-plane
magnetic fields. 
With respect to electromagnetic property, 
we will be able to regard the rippled graphene
with $\B_{\parallel}$ as a system of
traveling 2D massless Dirac fermions,
which are randomly scattered by ripples \cite{DSAHR11,LF10}.

In the present paper, we study a formula
of Lundeberg and Folk 
for the magnetoresistance $\Delta \rho$,
which was derived from a model based on
the Boltzmann transport theory \cite{LF10,LF10b}.
The relative height of ripple on the averaged 2D plane
is denoted by $h(\r)$ for each 2D position 
$\r=(x,y)$.
We characterize the height distribution
by the {\it two-point height correlation}
\begin{equation}
c(\r; \r_0)=\langle h(\r_0) h(\r_0+\r) \rangle,
\label{eqn:corr1}
\end{equation}
which is defined as a statistical average
of product of the height at $\r_0$ and that at the position 
separated from $\r_0$ by a displacement vector $\r$.
For simplicity, we assume that the distribution of $h(\r)$
is translation invariant and isotropic in this paper.
By this assumption, (\ref{eqn:corr1}) becomes
independent of $\r_0$ and it depends only on
the distance of two points $r=|\r|$;
$c(\r; \r_0)=c(r)$.
As a function of wave-number $k=|\k|$ of fermion,
we define 
\begin{equation}
g(k)=\int_0^{\infty} r W(rk) c(r) dr,
\label{eqn:g1}
\end{equation}
where
\begin{equation}
W(z)=\int_0^{2 \pi} J_0 \left( 2 z \sin \frac{\phi}{2} \right)
\sin^2 \frac{\phi}{2} d \phi
\label{eqn:W1}
\end{equation}
with the Bessel function $J_0(z)$ 
(see Appendix A for
special functions \cite{AAR99}).
The {\it signed carrier-density} of graphene is denoted by $n=n_+-n_-$,
which is the difference of
the density of electrons $n_+$
and that of holes $n_-$.
We assume that the temperature is low and the state
is degenerated so that
the wave-number of fermion $k$ is well-approximated
by the Fermi wave-number $k_{\rm F}$.
With counting
contributions from the two Dirac points
$\K$ and $\K'$ and spin,
$k \simeq k_{\rm F}$ is related with
$n$ as (see, for instance, Sect. II.A in \cite{CGPNG09}),
\begin{equation}
k=k(n)=\sqrt{\pi |n|},
\label{eqn:kn}
\end{equation}
where the absolute value of $n$ is 
used by the particle-hole symmetry;
$E_{\pm}=\pm \hbar v_{\rm F} k$.
Let $B_{\parallel}=|\B_{\parallel}|$ 
and $\theta$ be the angle
between $\B_{\parallel}$ and
the direction of the current $\J$ along which
the resistance $\rho$ is measured.
Set $\Delta \rho=\rho(B_{\parallel})-\rho(0)$,
the $\B_{\parallel}$-induced part of resistance, 
which we call {\it magnetoresistance}. 
Then the {\it Lundeberg-Folk formula} is given as
\begin{equation}
\Delta \rho(n, \theta, B_{\parallel})
=\frac{\pi B_{\parallel}^2}{2 \hbar}
(\sin^2 \theta +3 \cos^2 \theta) g(k(n)).
\label{eqn:LF1}
\end{equation}
Remark that this formula is not given
in the published paper \cite{LF10} of Lundeberg and Folk,
but the corresponding equations are found in
Appendix of its arXiv version \cite{LF10b}.
Lundeberg and Folk introduced a correlation length $R>0$
for ripple-height correlation $c(r)$ and argued
the asymptotic behavior of $g(k)$
in $k \ll 1/R$ and $k \gg 1/R$,
and from the latter estimate the behavior
of (\ref{eqn:LF1}) 
in high carrier-density region was predicted as
$\Delta \rho \sim |n|^{-3/2}, |n| \gg 1$.
This $|n|^{-3/2}$-law was emphasized in their paper
and has been used for analyzing experimental data
\cite{LF10,WS11,Shi12},
while it is an asymptotic law in $|n| \gg 1$
for a special case with the Gaussian height-correlation
of ripples as clarified below in the present paper.
Here we call the general equation (\ref{eqn:LF1})
the Lundeberg-Folk formula.

The main purpose of the present paper is 
to show that, for the two cases
\begin{eqnarray}
&& {\rm (i)} \quad c(r)=Z^2 e^{-r/R} 
\qquad \mbox{(exponential) \qquad and}
\nonumber\\
&&
{\rm (ii)} \quad c(r)=Z^2 e^{-(r/R)^2}
\qquad \mbox{(Gaussian)}
\nonumber
\end{eqnarray}
with a variance of height 
$Z^2=c(0)=\langle h(\r_0)^2 \rangle=\langle h^2 \rangle$, 
exact and explicit representations
for the anisotropic magnetoresistance (\ref{eqn:LF1})
are obtained.
(For the Gaussian correlated disorder, see \cite{Bar07}.)
The results are the following.
Let $F(\alpha, \beta, \gamma;z)$
and $F(\alpha, \gamma;z)$ be
the Gauss hypergeometric function and
the confluent hypergeometric function, respectively. 
(See Appendix A for definitions and basic properties
of special functions used in this paper.) 
Then, for (i) the exponential height-correlation,
(\ref{eqn:g1}) is calculated as
\begin{eqnarray}
g(k) &=& \pi (Z R)^2
F \left( \frac{3}{2}, \frac{3}{2}, 2; -(2k R)^2 \right)
\nonumber\\
&=& \frac{Z^2}{k^2} \Big[ K(2i k R)
-\frac{1}{1+(2k R)^2} E(2ikR) \Big],
\label{eqn:rhocase1}
\end{eqnarray}
where $K(z)$ and $E(z)$ are the complete elliptic integrals
of the first kind and of the second kind,
respectively (see Appendix A).
For (ii) the Gaussian height-correlation, 
(\ref{eqn:g1}) becomes
\begin{eqnarray}
g(k) &=& \frac{\pi Z^2 R^2}{2}
F \left( \frac{3}{2}, 2; - (k R)^2 \right)
\nonumber\\
&=& \frac{\pi (ZR)^2}{2}
\Big[ I_0( (k R)^2/2) - I_1( (k R)^2/2) \Big] e^{-(k R)^2/2},
\label{eqn:rhocase2}
\end{eqnarray}
where 
$I_{\nu}(z)$ is the modified Bessel function of the
first kind (see Appendix A).
From these exact expressions,
we can readily obtain the asymptotics of
the magnetoresistance (\ref{eqn:LF1}) in
high carrier-density region as follows. 
For $|n| \gg 1$,
\begin{eqnarray}
\Delta \rho(n, \theta, B_{\parallel}) \simeq
\left\{ \begin{array}{ll}
\displaystyle{
\frac{1}{2}(\sin^2 \theta+3 \cos^2 \theta)
\frac{(Z B_{\parallel})^2}{\sqrt{\pi} \hbar R}
|n|^{-3/2} \log(R |n|^{1/2})
},
&\mbox{for exponential (i)} \cr
& \cr
\displaystyle{
\frac{1}{4}(\sin^2 \theta+3 \cos^2 \theta)
\frac{(Z B_{\parallel})^2}{\hbar R}
|n|^{-3/2}
},
&\mbox{for Gaussian (ii)}.
\end{array} \right.
%\nonumber\\
\label{eqn:rhoasymA1}
\end{eqnarray}
The asymptotic for the Gaussian case (ii)
shown in the second line in (\ref{eqn:rhoasymA1})
is exactly the same as Eq.(3) reported in \cite{LF10}.
On the other hand, there is a logarithmic
correction to the $|n|^{-3/2}$-law in the case (i) with
the exponential height-correlation of ripples.

For the low carrier-density region,
we have found the following behavior for
the magnetoresistance;
when $|n| \simeq 0$, 
\begin{eqnarray}
\Delta \rho(n, \theta, B_{\parallel})
\simeq 
\left\{ \begin{array}{ll}
\displaystyle{\frac{1}{2}(\sin^2 \theta+3 \cos^2 \theta)
\frac{(\pi Z R B_{\parallel})^2}{\hbar}
\left(1- \frac{9}{2} \pi R^2 |n| \right)},
&\mbox{for exponential (i)} \cr
& \cr
\displaystyle{\frac{1}{4}(\sin^2 \theta+3 \cos^2 \theta)
\frac{(\pi Z R B_{\parallel})^2}{\hbar}
\left(1- \frac{3}{4} \pi R^2 |n| \right)},
&\mbox{for Gaussian (ii).}
\end{array} \right.
\label{eqn:rhoasymB1}
\end{eqnarray}
That is, $\rho(n, \theta, B_{\parallel})$ shows a cusp at
the charge neutrality point $|n|=0$
and the dependence on $n$ around this point is
linear $\propto -|n|$.

In the present paper we apply our analytic results
to the experimental data in the whole region of $n$.
The result seems to be excellent if we assume the
Gaussian height-correlation of ripples.
See Figures \ref{fig:Fig2} and \ref{fig:Fig3} 
in Section \ref{chap:remark}.
In the vicinity of charge neutrality point, 
electronic transport will be affected by 
strong density inhomogeneity in rippled graphene,
which is called {\it electron-hole puddle} \cite{DSAHR11}.
Moreover, in low carrier-density region, $|n| \ll 1$,
we have $k R \ll 1$ and the semi-classical
Boltzmann transport theory is afraid to be broken-down.
To the present analysis corrections by more elaborate approximations
or full quantum theory shall be added \cite{DSAHR11}.

The paper is organized as follows.
In Section \ref{chap:LF} we give a derivation of the
Lundeberg-Folk formula (\ref{eqn:LF1})
from a semi-classical model using the Boltzmann transport theory.
Section \ref{chap:proof} is devoted to the proofs of
our exact expressions (\ref{eqn:rhocase1}) and (\ref{eqn:rhocase2})
and their asymptotics (\ref{eqn:rhoasymA1}) and (\ref{eqn:rhoasymB1}).
In Section \ref{chap:remark}, we discuss comparison of the present
analytic results with the experimental data
measured by Wakabayashi and his coworkers.
Appendix \ref{chap:appendixA} is prepared 
for giving the mathematical formulas
for special functions used in the text.
In Appendix \ref{chap:appendixB} an explanation 
is given for the
linear relation between resistance
and inverse of relaxation time 
based on the Boltzmann transport theory.

%%%%%%%%%%%%%%%%%%%%%%%%%%%%%%%%%%%%%%%%%%%%%%%%%%
%%%  SEC2 %%%%%%%%%%%%%%%%%%%%%%%%%%%%%%%%%%%%%%%%
%%%%%%%%%%%%%%%%%%%%%%%%%%%%%%%%%%%%%%%%%%%%%%%%%%
\SSC{Derivation of Lundeberg-Folk formula}
\label{chap:LF}
%%%%%%%%%%%%%%%%%%%%%%%%%%%%%%%%%%%%%%%%%%%%%%%%%%

As remarked just below Eq.(\ref{eqn:LF1}),
the general expression of $\Delta \rho$
as a functional of the height-correlation function $c(r)$
of ripples, 
(\ref{eqn:LF1}) with (\ref{eqn:g1}) and (\ref{eqn:W1}),
is found only in \cite{LF10b}.
Although the essence of its derivation is given there,
the description is very brief.
For convenience, here we would like to complete
the derivation of the Lundeberg-Folk formula (\ref{eqn:LF1})
following \cite{LF10b}.

We take the averaged 2D plane of graphene as
the $x$-$y$ plane.
Let $h(\r)$ be the height of ripple at $\r=(x,y)$.

%%%%%%%%%%%%%%%%%%%%%%%%%%%%%%%%%
\subsection{Geometric vector potential}
%%%%%%%%%%%%%%%%%%%%%%%%%%%%%%%%%

Under the in-plane magnetic field in the $x$-direction,
\begin{equation}
\B_{\parallel}=(B_{\parallel}, 0),
\label{eqn:B1}
\end{equation}
the {\it geometric vector potential} of Berry \cite{Ber84},
$\A(\r)=(A_x(\r), A_y(\r))$, is calculated as \cite{MB01}
\begin{equation}
A_x(\r)=0, \qquad A_y(\r)=-B_{\parallel} h(\r).
\label{eqn:Berry1}
\end{equation}
It gives a potential expressed by 
a $2 \times 2$ complex matrix
for Dirac fermions with electric charge $-e$
and the Fermi velocity $v_{\rm F}$, 
\begin{eqnarray}
U(\r) &=& -e v_{\rm F} \A(\r) \cdot \vsigma
\nonumber\\
&=& e v_{\rm F} B_{\parallel} h(\r) \sigma_y
\nonumber\\
&=& e v_{\rm F} B_{\parallel} h(\r) 
\left( \begin{array}{cc}
0 & -i \cr
i & 0 
\end{array} \right), 
\label{eqn:U1}
\end{eqnarray}
where 
$\vsigma=(\sigma_x, \sigma_y)$ are Pauli matrices.

%%%%%%%%%%%%%%%%%%%%%%%%%%%%%%%%%
\subsection{Scattering matrix}
%%%%%%%%%%%%%%%%%%%%%%%%%%%%%%%%%

The scattering matrix for 
an interaction operator ${\cal U}$ 
is calculated by the
Fermi golden rule as
\begin{eqnarray}
S_{\rm ripple}(\k, \k') 
&=& \frac{2 \pi V}{\hbar}
|\langle \k'|{\cal U}|\k \rangle|^2
\delta(E(k')-E(k))
\nonumber\\
&=& \frac{2 \pi V}{\hbar^2 v_{\rm F}}
|\langle \k'|{\cal U}|\k \rangle|^2
\delta(k'-k),
\label{eqn:S1}
\end{eqnarray}
where $E(k)=\hbar v_{\rm F} k$.
We note that
\begin{equation}
\langle \k'|{\cal U}| \k \rangle
=\int d^2 r' \int d^2 r \,
\langle \k'|\r' \rangle
\langle \r'|{\cal U}|\r \rangle
\langle \r | \k \rangle
\label{eqn:U2}
\end{equation}
with
\begin{equation}
\langle \r' |{\cal U}| \r \rangle
=U(\r) \delta(\r-\r'), 
\label{eqn:U3}
\end{equation}
where the potential $U(\r)$ is given by (\ref{eqn:U1}),
$\delta(\r-\r')=\delta(x-x')\delta(y-y')$,
and $\langle \k'|\r \rangle = \langle \r' | \k' \rangle^{\dagger}$.
By using the two-component plane wave solution
(\ref{eqn:Dirac2}) for $E=E_+=\hbar v_{\rm F} k$,
(\ref{eqn:U2}) becomes
\begin{eqnarray}
\langle \k'|{\cal U}|\k \rangle
&=& \int d^2 r \, \frac{e^{-i \k' \cdot \r}}{\sqrt{2V}} 
(e^{i \theta_{k'}/2} \, e^{-i \theta_{k'}/2})
%\nonumber\\
%&& \times 
\left\{ e v_{\rm F} B_{\parallel} h(\r)
\left( \begin{array}{ll}
0 & -i \cr i & 0 
\end{array} \right) \right\}
\frac{e^{i \k \cdot \r}}{\sqrt{2V}}
\left( \begin{array}{l}
e^{-i \theta_k/2} \cr e^{i \theta_k/2}
\end{array} \right)
\nonumber\\
&=& \frac{e v_{\rm F} B_{\parallel}}{V}
\sin \frac{\theta_{k}+\theta_{k'}}{2} \widehat{h}^{*}(\q)
\label{eqn:U4}
\end{eqnarray}
with $\q=\k'-\k$ and
\begin{equation}
\widehat{h}^{*}(\q)=\int d^2 r \, e^{-i \q \cdot \r} h(\r).
\label{eqn:hq1}
\end{equation}
The square of (\ref{eqn:hq1}), 
$|\widehat{h}(\q)|^2=\widehat{h}(\q) \widehat{h}^*(\q)$, is
then calculated as
\begin{eqnarray}
|\widehat{h}(\q)|^2 &=& \int d^2 r \, \int d^2 r' \, e^{i \q \cdot (\r'-\r)}
h(\r) h(\r')
\nonumber\\
&=& V \int d^2 r \, e^{i \q \cdot \r}
\frac{1}{V} \int d^2 r_0 \, h(\r_0) h(\r_0+\r),
\label{eqn:hq2}
\end{eqnarray}
where we have renamed the integration variable
as $\r \to \r_0$ and $\r'-\r \to \r$.
The quantity involving the second integral in 
the second line of (\ref{eqn:hq2}),
$(1/V) \int d^2 r_0 h(\r_0) h(\r_0+\r)$,
is a volume average of a product $h(\r_0) h(\r_0+\r)$
with respect to $\r_0$ over the system.
If the distribution of the ripple height $h(\r)$ on graphene
is translation invariant, this volume average
will converge to the height correlation function (\ref{eqn:corr1})
as $V \to \infty$,
by the self-averaging property of usual random systems.
Moreover, it can be written as $c(r)$, if the system is also isotropic.
Therefore, provided that $V$ is large enough,
(\ref{eqn:hq2}) is equal to
the Fourier transform of $c(r)$
\begin{equation}
\widehat{c}(q)=\int d^2 r \, 
e^{i \q \cdot \r} c(r),
\quad q= |\q|
\label{eqn:cq1}
\end{equation}
multiplied by the volume $V$ of the system;
$|\widehat{h}(\q)|^2=V \widehat{c}(q)$.
Inserting the above results into (\ref{eqn:S1}),
the volume factor $V$ is cancelled out as desired, and
we obtain the following expression of scattering matrix, 
\begin{equation}
S_{\rm ripple}(\k, \k')=\frac{2 \pi e^2 v_{\rm F} B_{\parallel}^2}{\hbar^2}
\sin^2 \frac{\theta_{k}+\theta_{k'}}{2}
\widehat{c}(q) \delta(k'-k),
\quad \q=\k'-\k, \quad q=|\q|,
\label{eqn:S2}
\end{equation}
which was given as Eq.(A.1) in \cite{LF10b}.

By definition of $\theta_k$ for $\k=(k_x, k_y)$
given by (\ref{eqn:Dirac3}) and its analogue 
$\theta_{k'}$ for $\k'=(k_x', k_y')$, 
under the energy conservation $k'=k$,
$q_x=k_x'-k_x=k(\cos \theta_{k'}-\cos \theta_k)$,
$q_y=k_y'-k_y=k(\sin \theta_{k'}-\sin \theta_k)$,
and thus
\begin{eqnarray}
q &=& |\q| = \sqrt{q_x^2+q_y^2}
\nonumber\\
&=& 2k \sqrt{1-\cos(\theta_{k}-\theta_{k'})}
\nonumber\\
&=& 2 k \sin \frac{\theta_k-\theta_{k'}}{2}.
\label{eqn:q1}
\end{eqnarray}
Then, when $\q=\k'-\k$ with the condition $k'=k$,
(\ref{eqn:cq1}) becomes
\begin{eqnarray}
\widehat{c}(q) &=& \int_0^{\infty} dr \, r
\int_{0}^{2 \pi} d \varphi \, c(r) e^{i q r \cos \varphi}
\nonumber\\
&=& \int_0^{\infty} dr \, r c(r) 
\int_0^{2 \pi} d\varphi 
\exp \left[ 2 i k r \sin \frac{\theta_k-\theta_{k'}}{2}
\cos \varphi \right].
\label{eqn:cq2}
\end{eqnarray}

%%%%%%%%%%%%%%%%%%%%%%%%%%%%%%%%%
\subsection{Boltzmann equation and scattering rates}
%%%%%%%%%%%%%%%%%%%%%%%%%%%%%%%%%

Lundeberg and Folk introduced a semi-classical model
based on the Boltzmann transport theory
for the system, in which the 2D electronic state with 
a wave-number vector $\k$ is described by
probability distribution $f(\k)$.
The time evolution of $f(\k)$ follows
the Boltzmann equation
(see, for example, \cite{Kubo91,BMB04})
\begin{equation}
\left( \frac{\partial}{\partial t}
+\frac{e}{\hbar} \E \cdot \nabla_{k} \right) f(\k)
=\left[ \frac{\partial f(\k)}{\partial t} \right]_{\rm S},
\label{eqn:Boltzmann0}
\end{equation}
in the external electric field $\E$,
where we have assumed the spatial homogeneity
of distribution $f$ and omitted the term
$\v \cdot \nabla_{r} f(\k)$ in the LHS.
The `collision term' in the RHS, which 
describes the scattering of particles or holes
by random vector potentials caused by ripples
in the present system, is given by
\begin{equation}
\left[\frac{\partial  f(\k)}{\partial t} \right]_{\rm S}
=D[S, f](\k)
\label{eqn:Boltzmann}
\end{equation}
with the operator
\begin{equation}
D[S, f](\k)
=\int \frac{d^2 k'}{(2\pi)^2}
S(\k, \k') \Big[ f(\k')-f(\k) \Big]
\label{eqn:D1}
\end{equation}
where 
$S=S_0+S_{\rm ripple}$ is the total scattering matrix.
Here $S_0$ is the zero field scattering matrix assumed to be
in the isotropic form,
$S_0(\k, \k')=s_{0}(k,q) \delta(k-k')$, 
and $S_{\rm ripple}$ is given by (\ref{eqn:S2}).
We set $S(\k, \k')=\widetilde{S}(\k, \k') \delta(k-k')$
and rewrite (\ref{eqn:D1}) as 
\begin{equation}
D[S, f](\k)
=\frac{k}{(2\pi)^2}
\int_{-\pi}^{\pi} d \theta_{k'} \,
\widetilde{S}(\k, \k')
\Big[ \widetilde{f}(\theta_{k'})
-\widetilde{f}(\theta_k) \Big],
\label{eqn:D2}
\end{equation}
where $\widetilde{f}(\theta_k)$
shows the dependence of $f(\k)$ on the 
angular component $\theta_k$ of $\k$,
when $k=|\k|$ is given.
In order to study $\theta_k$-dependence
of scattering induced by $\B_{\parallel}$, 
which is imposed in the $x$-direction,
we introduce the following orthogonal bases,
\begin{equation}
\widetilde{f}_{x}(\theta_k)=\frac{\cos \theta_k}{\sqrt{\pi}},
\quad
\widetilde{f}_{y}(\theta_k)=\frac{\sin \theta_k}{\sqrt{\pi}},
\label{eqn:f1}
\end{equation}
satisfying
\begin{equation}
\int_{-\pi}^{\pi} d \theta_k \,
\widetilde{f}_{\alpha}(\theta_k) \widetilde{f}_{\beta}(\theta_k)
=\delta_{\alpha, \beta},
\qquad \alpha, \beta=x,y.
\label{eqn:orth}
\end{equation}
The $\B_{\parallel}$-induced parts of scattering rates, 
which give the inverses of
transport times enhanced by $\B_{\parallel}$, 
are obtained as
\begin{equation}
\Delta \tau_{\alpha, \beta}^{-1}(k)
=-\int_{-\pi}^{\pi} d \theta_{k} \,
\widetilde{f}_{\alpha}(\theta_{k})
D[\widetilde{S}_{\rm ripple}, \widetilde{f}_{\beta}](\theta_k)
\label{eqn:tau1}
\end{equation}
for $\alpha, \beta=x, y$,
by the orthogonality relation (\ref{eqn:orth}) 
of the two bases (\ref{eqn:f1}).

Now combining (\ref{eqn:tau1}) and (\ref{eqn:D2})
with (\ref{eqn:S2}) and (\ref{eqn:cq2}),
we have multiple-integral expressions
for $\Delta \tau_{\alpha, \beta}^{-1}$ 
such that integrals are performed
with respect to $\theta_k, \theta_{k'}, r$ and $\varphi$.
Change the integral variables as
$\theta_k+\theta_{k'} \to \theta_+$,
$\theta_k-\theta_{k'} \to \theta_-$,
and let
\begin{eqnarray}
L_{\alpha, \beta}(k, r, \theta_-)
&=& \int_0^{4 \pi} d \theta_+ 
\int_{0}^{2 \pi} d \varphi 
\exp \left[ 2 i k r \sin \frac{\theta_-}{2}
\cos \varphi \right] \sin^2 \frac{\theta_+}{2}
\nonumber\\
&& \times \widetilde{f}_{\alpha}\left( \frac{\theta_+ + \theta_-}{2} \right)
\left[ \widetilde{f}_{\beta}\left( \frac{\theta_+ - \theta_-}{2} \right)
-\widetilde{f}_{\beta} \left( \frac{\theta_+ + \theta_-}{2} \right) \right],
\label{eqn:L1}
\end{eqnarray}
$\alpha, \beta = x, y$. Then
\begin{equation}
\Delta \tau_{\alpha, \beta}^{-1}(k)
=- \frac{e^2 v_{\rm F} B_{\parallel}^2}
{4 \pi \hbar^2} k \int_0^{\infty} dr \, r c(r)
\int_0^{2 \pi} d \theta_{-} \,
L_{\alpha, \beta}(k, r, \theta_-),
\quad \alpha, \beta = x, y.
\label{eqn:tau2}
\end{equation}
By the Boltzmann transport theory
(see Appendix B for an explanation),
the resistances induced by $\B_{\parallel}$
are related with $\tau_{\alpha, \beta}^{-1}$ as
\begin{equation}
\Delta \rho_{\alpha, \beta}(k)
=\frac{2 \pi \hbar}{e^2 v_{\rm F} k}
\Delta \tau_{\alpha, \beta}^{-1}(k),
\quad \alpha, \beta = x, y.
\label{eqn:rhoB1}
\end{equation}

We performed the integrals (\ref{eqn:L1})
and obtained the results
\begin{eqnarray}
\label{eqn:L2a}
&& L_{xx}(k, r, \theta_-)
= 3 L_{yy}(k, r, \theta_-)
=-3 \pi J_{0} \left( 2 k r \sin \frac{\theta_-}{2} \right)
\sin^2 \frac{\theta_-}{2},
\\
\label{eqn:L2b}
&& L_{xy}(k, r, \theta_-)
= - \frac{1}{3} L_{yx}(k, r, \theta_-)
=- \pi J_{0} \left( 2 k r \sin \frac{\theta_-}{2} \right)
\sin \theta_-,
\end{eqnarray}
where $J_0(z)$ is the Bessel function with index $\nu=0$
(see Appendix A).

Then (\ref{eqn:rhoB1}) with (\ref{eqn:tau2}) gives
\begin{eqnarray}
\label{eqn:rhoB2a}
&& \Delta \rho_{xx}(k) = 3 \Delta \rho_{yy}(k)
=\frac{3 \pi B_{\parallel}^2}{2 \hbar} g(k),
\\
\label{eqn:rhoB2b}
&& \Delta \rho_{xy}(k)=\Delta \rho_{yx}(k)=0,
\end{eqnarray}
where $g(k)$ is given by (\ref{eqn:g1})
with (\ref{eqn:W1}).
The `threefold relation'
between $\Delta \rho_{xx}$ and $\Delta \rho_{yy}$
is found in (\ref{eqn:rhoB2a}) \cite{Rushforth04,LF10,LF10b}.

%%%%%%%%%%%%%%%%%%%%%%%%%%%%%%%%%
\subsection{Lundeberg-Folk formula}
%%%%%%%%%%%%%%%%%%%%%%%%%%%%%%%%%

Let $\E=(E_x, E_y)$ be the in-plane electric field
applied in order to measure resistance.
If the current is written as $\J=(J_x, J_y)$, 
then the resistance coefficients $\Delta \rho_{\alpha, \beta}$,
$\alpha, \beta=x,y$, 
are defined as
\begin{equation}
\left( \begin{array}{l} E_x \cr E_y \end{array} \right)
= \left( \begin{array}{ll} \Delta \rho_{xx} & \Delta \rho_{xy} \cr
\Delta \rho_{yx} & \Delta \rho_{yy} \end{array} \right)
\left( \begin{array}{l} J_x \cr J_y \end{array} \right).
\label{eqn:EJ1}
\end{equation}
In the isotropic system, 
$\Delta \rho_{xy} = \Delta \rho_{yx}=0$ as in (\ref{eqn:rhoB2b}),
then $E_x=\Delta \rho_{xx} J_x$,
$E_y=\Delta \rho_{yy} J_y$.
We have chosen the direction of $\B_{\parallel}$
in the $x$-direction as (\ref{eqn:B1}).
Then, if we denote the angle between $\B_{\parallel}$
and $\J$ as $\theta$,
$J_x=J \cos \theta, J_y=J \sin \theta$
with $J=|\J|$ and
the $\B_{\parallel}$-induced resistance $\Delta \rho$
for the current $\J$ is given by
\begin{eqnarray}
\Delta \rho &=& \frac{\E \cdot \J}{J^2}
\nonumber\\
&=& \frac{1}{J^2} (\Delta \rho_{xx} J_x^2
+ \Delta \rho_{yy} J_y^2)
\nonumber\\
&=& \Delta \rho_{xx} \cos^2 \theta+\Delta \rho_{yy} \sin^2 \theta.
\label{eqn:rhoC1}
\end{eqnarray}
Then by the threefold relation (\ref{eqn:rhoB2a}) 
between $\Delta \rho_{xx}$ and $\Delta \rho_{yy}$, 
the Lundeberg-Folk formula
(\ref{eqn:LF1}) is obtained.

%%%%%%%%%%%%%%%%%%%%%%%%%%%%%%%%%%%%%%%%%%%%%%%%%%
%%%  SEC3 %%%%%%%%%%%%%%%%%%%%%%%%%%%%%%%%%%%%%%%%
%%%%%%%%%%%%%%%%%%%%%%%%%%%%%%%%%%%%%%%%%%%%%%%%%%
\SSC{Proofs of results}
\label{chap:proof}
%%%%%%%%%%%%%%%%%%%%%%%%%%%%%%%%%%%%%%%%%%%%%%%%%%

%%%%%%%%%%%%%%%%%%%%%%%%%%%%%%%%%%%%%%%%%
\subsection{Proof of Eq.(\ref{eqn:rhocase1})}
%%%%%%%%%%%%%%%%%%%%%%%%%%%%%%%%%%%%%%%%%

Let $c(r)=Z^2 e^{-r/R}$ and use the formula (\ref{eqn:Jnu1})
for $\nu=0$. Then (\ref{eqn:g1}) with (\ref{eqn:W1}) 
is written as
\begin{equation}
g(k)=Z^2 \sum_{n=0}^{\infty} \frac{(-k^2)^n}{(n!)^2}
M_r(n) M_{\phi}(n)
\label{eqn:g_A1}
\end{equation}
with
\begin{eqnarray}
M_r(n) &=& \int_0^{\infty} dr \, r^{2n+1} e^{-r/R}
= R^{2n+2} \Gamma(2n+2),
\nonumber\\
M_{\phi}(n) &=& \int_0^{2 \pi} d \phi \,
\sin^{2n+2} \frac{\phi}{2}
=\frac{2 \sqrt{\pi} \Gamma(n+3/2)}{\Gamma(n+2)}.
\label{eqn:Ms}
\end{eqnarray}
We apply the duplication formula (\ref{eqn:Gamma2}) of Gamma functions
and use the Pochhammer symbol (\ref{eqn:Poch}),
then we have
\begin{eqnarray}
g(k) &=& (ZR)^2 \sum_{n=0}^{\infty}
\frac{\{2 \Gamma(n+3/2)\}^2}{n! \Gamma(n+2)}
(-(2kR)^2)^n
\nonumber\\
&=& \pi (ZR)^2 \sum_{n=0}^{\infty} 
\frac{\{(3/2)_n\}^2}{(2)_n}
\frac{(-(2kR)^2)^n}{n!},
\label{eqn:g_A2}
\end{eqnarray}
which prove the first equality of (\ref{eqn:rhocase1})
by the definition of the Gauss hypergeometric function
(\ref{eqn:F1}).

From the recurrence relations (\ref{eqn:G_rec1})
and (\ref{eqn:G_rec2}), we obtain the following
two equalities,
\begin{eqnarray}
&& F \left( \frac{3}{2}, \frac{3}{2}, 2; -(2kR)^2 \right)
=\frac{1}{2 (kR)^2}
\left[ F \left( \frac{1}{2}, \frac{1}{2},1; -(2kR)^2 \right)
- F \left( \frac{1}{2}, \frac{3}{2},1; -(2kR)^2 \right) \right],
\nonumber\\
&&  F \left( \frac{1}{2}, \frac{3}{2},1; -(2kR)^2 \right)
=\frac{1}{1+(2kR)^2} 
 F \left( -\frac{1}{2}, \frac{1}{2},1; -(2kR)^2 \right).
\label{eqn:F1rec1}
\end{eqnarray}
Combining them gives 
\begin{eqnarray}
&&F \left( \frac{3}{2}, \frac{3}{2},2; -(2kR)^2 \right)
\nonumber\\
&& \quad =\frac{1}{2(kR)^2} 
\left[  F \left( \frac{1}{2}, \frac{1}{2},1; -(2kR)^2 \right)
- \frac{1}{1+(2kR)^2}
 F\left( -\frac{1}{2}, \frac{1}{2},1; -(2kR)^2 \right) \right].
\label{eqn:F1rec2}
\end{eqnarray}
By the formulas (\ref{eqn:KE2}), which express
the complete elliptic integrals $K(z)$ and $E(z)$
by using the Gauss hypergeometric functions,
the second equality of (\ref{eqn:rhocase1}) is proved.

%%%%%%%%%%%%%%%%%%%%%%%%%%%%%%%%%%%%%%%%%
\subsection{Proof of Eq.(\ref{eqn:rhocase2})}
%%%%%%%%%%%%%%%%%%%%%%%%%%%%%%%%%%%%%%%%%

Let $c(r)=Z^2 e^{- (r/R)^2}$ and use the formula (\ref{eqn:Jnu1})
for $\nu=0$. Then (\ref{eqn:g1}) with (\ref{eqn:W1}) 
is written as
\begin{equation}
g(k)=Z^2 \sum_{n=0}^{\infty} \frac{(-k^2)^n}{(n!)^2}
\widetilde{M}_r(n) M_{\phi}(n), 
\label{eqn:g_B1}
\end{equation}
where
\begin{equation}
\widetilde{M}_r(n) = \int_0^{\infty} dr \, r^{2n+1} e^{-r^2/R^2}
= \frac{1}{2} R^{2n+2} n!
\label{eqn:Ms2}
\end{equation}
and $M_{\phi}(n)$ is given as in (\ref{eqn:Ms}).
Following the similar procedure mentioned above, 
we have
\begin{equation}
g(k) = \frac{\pi (ZR)^2}{2} \sum_{n=0}^{\infty} 
\frac{(3/2)_n}{(2)_n}
\frac{(-(kR)^2)^n}{n!},
\label{eqn:g_B2}
\end{equation}
which proves the first equality of (\ref{eqn:rhocase2})
by the definition of the confluent hypergeometric function
(\ref{eqn:F2}).

From the recurrence relations (\ref{eqn:c_rec1})
and (\ref{eqn:c_rec2}), we obtain the following equality,
\begin{equation}
F(\alpha+1, \gamma;z)=F(\alpha, \gamma-1;z)
+\frac{z(\alpha-\gamma+1)}{\gamma(1-\gamma)} 
F(\alpha+1, \gamma+1;z).
\label{eqn:F2rec}
\end{equation}
By using this equality for $\alpha=1/2$ and $\gamma=2$,
we can rewrite the first line of (\ref{eqn:rhocase2}) as
\begin{equation}
g(k)= \frac{\pi (ZR)^2}{2}
\left[ F \left( \frac{1}{2}, 1; -(kR)^2 \right)
-\frac{(kR)^2}{4} 
F \left( \frac{3}{2}, 3; -(kR)^2 \right) \right].
\label{eqn:g_B3}
\end{equation}
By the formula (\ref{eqn:Inu2}), which expresses
the modified Bessel function $I_{\nu}(z)$ 
by using the confluent hypergeometric functions,
the second equality of (\ref{eqn:rhocase2}) is proved.

%%%%%%%%%%%%%%%%%%%%%%%%%%%%%%%%%%%%%%%%%
\subsection{Asymptotics in high carrier-density region}
%%%%%%%%%%%%%%%%%%%%%%%%%%%%%%%%%%%%%%%%%

\noindent{\it Case (i): the exponential height-correlation}
\vskip 0.3cm

In Gauss' integral representation (\ref{eqn:GaussInt}),
if we change the variable as
$u=(w-1)/(4(kR)^2)$, we have
\begin{equation}
F \left( \frac{3}{2}, \frac{3}{2}, 2; -(2kR)^2 \right)
=\frac{1}{4 \pi (kR)^3}
\int_1^{1+4(kR)^2}
dw \, (w-1)^{1/2} \left(1-\frac{w-1}{4(kR)^2} \right)^{-1/2}
w^{-3/2}.
\label{eqn:asymF1z}
\end{equation}
Then we obtain the equality
\begin{eqnarray}
&& F \left( \frac{3}{2}, \frac{3}{2}, 2; -(2kR)^2 \right)
\nonumber\\
&& 
= \frac{1}{\pi (kR)^3}
\left( 4 + \frac{1}{(kR)^2} \right)^{-1/2}
\left[ K \left( \sqrt{\frac{(2kR)^2}{1+(2kR)^2}} \right)
- E \left( \sqrt{\frac{(2kR)^2}{1+(2kR)^2}} \right) \right].
\label{eqn:asymF1}
\end{eqnarray}
By the asymptotics of the complete elliptic integrals
\cite{AS72}
\begin{eqnarray}
K(z) &\simeq& \log \left( \frac{4}{\sqrt{1-z^2}} \right),
\nonumber\\
E(z) &\simeq& 1,
\qquad \mbox{as $z \to 1$},
\label{eqn:asymKE1}
\end{eqnarray}
we have the behavior
\begin{equation}
F \left( \frac{3}{2}, \frac{3}{2}, 2; -(2kR)^2 \right)
\simeq \frac{1}{\pi (kR)^2}
\Big[ \log(4 \sqrt{1+(2kR)^2}) -1 \Big],
\quad \mbox{as $k \to \infty$}.
\label{eqn:asymF2}
\end{equation}
Through the relation (\ref{eqn:kn}) between $k$ and $n$,
the Lundeberg-Folk formula (\ref{eqn:LF1})
with the present result (\ref{eqn:rhocase1}) gives
the first line of the RHS of (\ref{eqn:rhoasymA1}). 

\vskip 0.3cm
\noindent{\it Case (ii): the Gaussian height-correlation}
\vskip 0.3cm

In this case, we use the asymptotic expansion 
formula of the confluent hypergeometric function,
\begin{eqnarray}
F(\alpha, \gamma; z)
&\simeq& \frac{\Gamma(\gamma)}{\Gamma(\gamma-\alpha)}
(-z)^{-\alpha}
\sum_{n=0}^{\infty}(-1)^n
\frac{(\alpha)_n (\gamma-\alpha)_n}{n!} z^{-n}
\nonumber\\
&& + \frac{\Gamma(\gamma)}{\Gamma(\alpha)}
e^{z} z^{\alpha-\gamma}
\sum_{n=0}^{\infty} \frac{(1-\alpha)_n (\gamma-\alpha)_n}{n!}
z^{-n}.
\label{eqn:asymexp}
\end{eqnarray}
It gives
\begin{equation}
F \left( \frac{3}{2}, 2; -(kR)^2 \right)
\simeq \frac{1}{\sqrt{\pi} (kR)^3}
+{\cal O}((kR)^{-5}),
\label{eqn:asymexp2}
\end{equation}
and we obtain the second line of the RHS of
(\ref{eqn:rhoasymA1}).
This result is exactly the same as Eq.(3)
reported in \cite{LF10}.

%%%%%%%%%%%%%%%%%%%%%%%%%%%%%%%%%%%%%%%%%
\subsection{Behavior in low carrier-density region}
%%%%%%%%%%%%%%%%%%%%%%%%%%%%%%%%%%%%%%%%%

\noindent{\it Case (i): the exponential height-correlation}
\vskip 0.3cm

By using the first two terms of the series (\ref{eqn:F1})
defining the Gauss hypergeometric function,
we obtain from (\ref{eqn:rhocase1})
\begin{equation}
g(k)=\pi (ZR)^2 \left[
1-\frac{9}{2}(kR)^2+{\cal O}((kR)^4) \right].
\label{eqn:g_exp1}
\end{equation}
The first line of the RHS of (\ref{eqn:rhoasymB1}) 
is concluded from (\ref{eqn:g_exp1}).

\vskip 0.3cm
\noindent{\it Case (ii): the Gaussian height-correlation}
\vskip 0.3cm

By using the first two terms of the series (\ref{eqn:F2})
defining the confluent hypergeometric function,
we obtain from (\ref{eqn:rhocase2})
\begin{equation}
g(k)= \frac{\pi (ZR)^2}{2} \left[
1-\frac{3}{4}(kR)^2+{\cal O}((kR)^4) \right].
\label{eqn:g_exp2}
\end{equation}
It gives the second line of the RHS of
(\ref{eqn:rhoasymB1}).

%%%%%%%%%%%%%%%%%%%%%%%%%%%%%%%%%%%%%%%%%%%%%%%%%%
%%%  SEC4 %%%%%%%%%%%%%%%%%%%%%%%%%%%%%%%%%%%%%%%%
%%%%%%%%%%%%%%%%%%%%%%%%%%%%%%%%%%%%%%%%%%%%%%%%%%
\SSC{Concluding remarks}
\label{chap:remark}
%%%%%%%%%%%%%%%%%%%%%%%%%%%%%%%%%%%%%%%%%%%%%%%%%%

The Lundeberg-Folk formula (\ref{eqn:LF1})
shows that the magnetoresistance induced by
$\B_{\parallel}$ is proportional to $B_{\parallel}^2$,
\begin{equation}
\Delta \rho(n, \theta, B_{\parallel})
= a(n, \theta) B_{\parallel}^2.
\label{eqn:LFB1}
\end{equation}
The coefficient $a(n, \theta)$ is given by
\begin{equation}
a(n, \theta)=\frac{\pi}{2 \hbar}
(\sin^2 \theta+ 3 \cos^2 \theta) g(\sqrt{\pi |n|}),
\label{eqn:LFB2}
\end{equation}
in which the function $g$ depends on the height-correlation
function $c(r)$ of ripples on graphene.
In the present paper, we have clarified the dependence
of $\Delta \rho$ on $c(r)$ for the two cases,
(i) $c(r)=Z^2 e^{-r/R}$ (exponential)
and (ii) $c(r)=Z^2 e^{-(r/R)^2}$ (Gaussian),
where $Z^2=c(0)=\langle h^2 \rangle$
is the variance of height of ripples
and $R$ indicates the correlation length of ripples.

In the low carrier-density region, the so-called
electron-hole puddles appear on the rippled graphene
and the homogeneity assumption of system will be invalid.
In order to handle the inhomogeneity of the carrier density
around the Dirac point, the present Boltzmann approach
should be corrected including higher-level
approximations \cite{DSAHR11,RDS08,Rossi09}.
Away from the charge neutrality point, however,
the semi-classical modeling based on the Boltzmann transport theory
works well to describe the electronic transport properties
of graphene.
Here we consider our analytic expressions for $\Delta \rho$
as `trial interpolation formulas'
connecting two high-carrier-density regimes, $|n| \gg 1$;
the positive-charge regime ($n>0$) and the negative one ($n < 0$).

Our expressions for $\Delta \rho$
reported in this paper include
two parameters $Z$ and $R$.
If experimental data are provided,
we can evaluate 
the standard deviation of ripple height 
$Z=\sqrt{\langle h^2 \rangle}$ and 
the correlation length of ripples $R$
by numerical fitting of the data
to our analytic expressions (\ref{eqn:rhocase1})
or (\ref{eqn:rhocase2}).

%%%%%%%%%%%%%%% Figure 1 %%%%%%%%%%%%%%%%%%%%%%%%%%%%%%%%%%%%%%%%%
\begin{figure}
\includegraphics[width=1.0\linewidth]{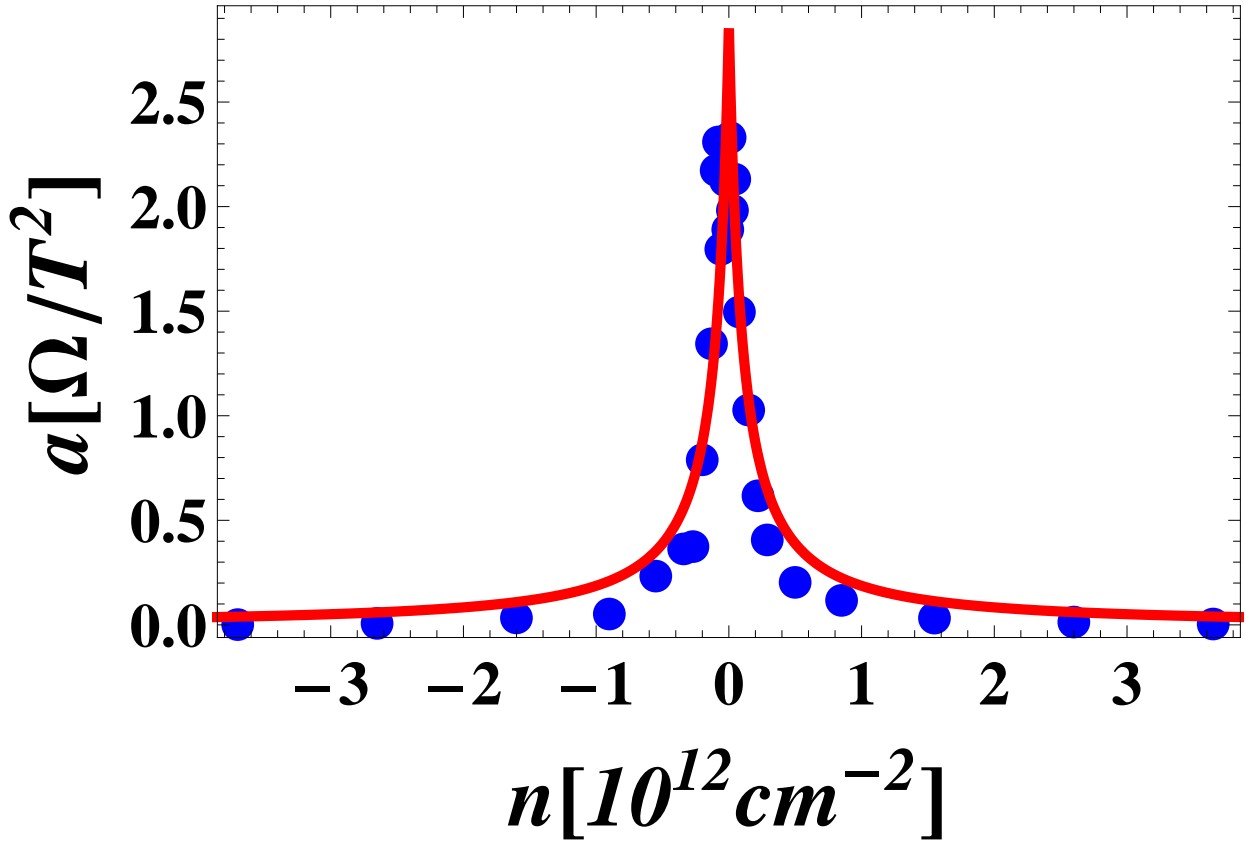}
\caption{
The coefficient $a(n, \theta)$ of the
Lundeberg-Folk formula (\ref{eqn:LFB2}) 
for $\theta=0$ with the present
exact solution (\ref{eqn:rhocase1}) for the
exponential height-correlation of ripples
is drawn. The parameters are
chosen as $Z=0.531$ [nm] and $R=8.51$ [nm]
by the best fitting to the experimental data given 
by Wakabayashi and his coworkers 
(indicated by dots).
\label{fig:Fig1}}
\end{figure}
%%%%%%%%%%%%%%%%%%%%%%%%%%%%%%%%%%%%%%%%%%%%%%%%%%%%%%%%%%%%%%%%%%

%%%%%%%%%%%%%%% Figure 2 %%%%%%%%%%%%%%%%%%%%%%%%%%%%%%%%%%%%%%%%%
\begin{figure}
\includegraphics[width=1.0\linewidth]{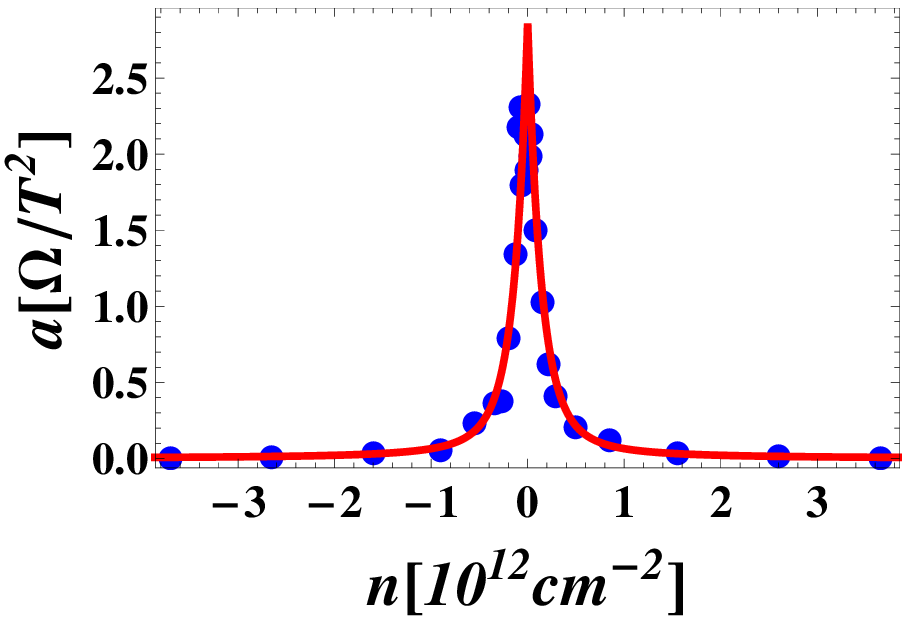}
\caption{
The coefficient $a(n, \theta)$ of the
Lundeberg-Folk formula (\ref{eqn:LFB2}) 
for $\theta=0$ with the present
exact solution (\ref{eqn:rhocase2}) for the
Gaussian height-correlation of ripples
is drawn. The parameters are
chosen as $Z=0.376$ [nm] and $R=17.0$ [nm]
by the best fitting to the experimental data given 
by Wakabayashi and his coworkers 
(indicated by dots).
The curve fits the experimental data very well
in the whole region of carrier-density $n$.
\label{fig:Fig2}}
\end{figure}
%%%%%%%%%%%%%%%%%%%%%%%%%%%%%%%%%%%%%%%%%%%%%%%%%%%%%%%%%%%%%%%%%%

Figures \ref{fig:Fig1} and \ref{fig:Fig2}
show results of numerical fitting
of the Lundeberg-Folk formula (\ref{eqn:LF1})
with our exact solutions 
(\ref{eqn:rhocase1}) and (\ref{eqn:rhocase2})
for the exponential and the Gaussian height-correlations
to the experimental data, respectively, 
when $\theta=0$ in the both cases. 
The experimental data are given by
Wakabayashi and his coworkers \cite{WS11,Shi12}.
As we can see in these figures, 
the fitting in the high carrier-density region
is more excellent in the Gaussian case 
(Fig.\ref{fig:Fig2}) than in the exponential case
(Fig.\ref{fig:Fig1}). 

%%%%%%%%%%%%%%% Figure 3 %%%%%%%%%%%%%%%%%%%%%%%%%%%%%%%%%%%%%%%%%
\begin{figure}
\includegraphics[width=1.0\linewidth]{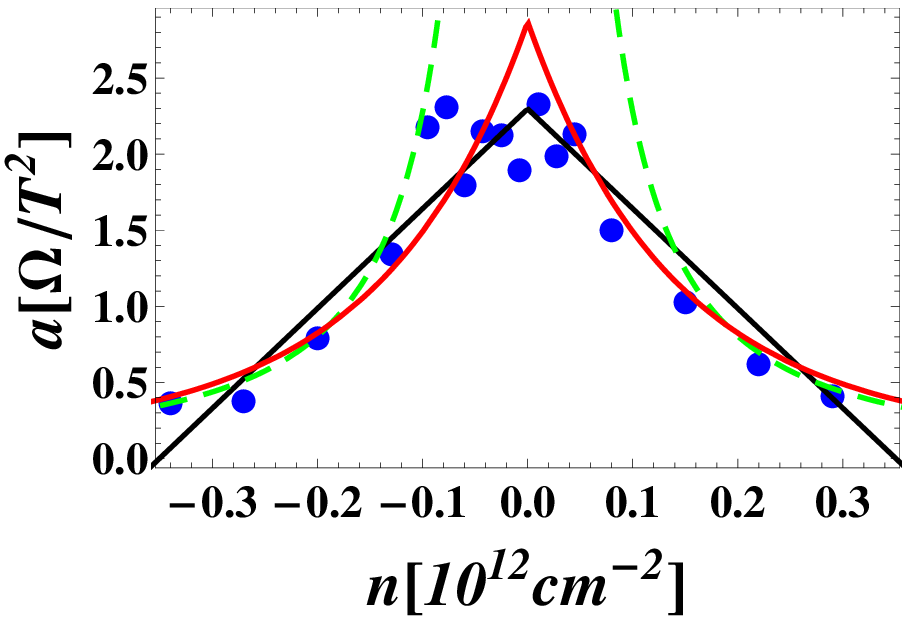}
\caption{Enlarged figure of Fig.\ref{fig:Fig2} 
in the vicinity of the
charge neutrality point ($|n|=0$).
Dots are the experimental data given by
Wakabayashi and his coworkers.
The bold curve showing a cusp at $|n|=0$
is the present analytic solution
(\ref{eqn:rhocase2}) for the Gaussian height-correlation
with $Z=0.376$ [nm] and $R=17.0$ [nm].
In this low carrier-density region,
the $|n|^{-3/2}$-law shown by dotted curves
is invalid, since it diverges as $|n| \to 0$.
In the very vicinity of the point $|n|=0$,
the simple expression given in
the second line of the RHS of (\ref{eqn:rhoasymB1})
will work well, which is shown by 
thin black lines.
\label{fig:Fig3}}
\end{figure}
%%%%%%%%%%%%%%%%%%%%%%%%%%%%%%%%%%%%%%%%%%%%%%%%%%%%%%%%%%%%%%%%%%

For the vicinity of the charge neutrality point,
$|n| \simeq 0$,
the fitting result for the Gaussian height-correlation
is shown by Fig. \ref{fig:Fig3},
in which the $|n|^{-3/2}$-law for the
high carrier-density region $|n| \gg 1$
(the second line of (\ref{eqn:rhoasymA1}))
and the expression for $|n| \simeq 0$
(the second line of (\ref{eqn:rhoasymB1}))
are also plotted for comparison.
The obtained values of parameters by the present
numerical fitting 
are $Z=0.376$ [nm] and $R=17.0$ [nm],
respectively.
These values should be compared with
the direct experimental measurements
\cite{ICCFW07,Geringer09,LF10,WS11,Shi12}.
In particular, it was reported in \cite{Shi12}
the AFM measurements gave 
$Z=0.118$ [nm] and $R=15.2$ [nm].
See also experimental values cited in \cite{LF10,LF10b}
from \cite{ICCFW07,Geringer09} and others.
(We note that recent high-resolution measurements
and analysis have revealed smaller-sized roughness
of the substrate and graphene surfaces \cite{Cullen10}.)

We hope that the our expressions will be
used for systematic analysis of experimental data
also for different values of $\theta$.
In the present paper, we have assumed translation
invariance and isotropy for distribution
of ripple height.
Even in the high carrier-density regions,
apart from the effect of electron-hole puddles, 
further calculation for general setting 
of ripple-height-correlations will be
an interesting future problem.

%%%%%%%%%%%%%%%%%%%%%%%%%%%%%%%%%%%%%%%%%%%%%%%%
\vskip 1cm
%%%%%%%%%%%%%%%%%%%%%%%%%%%%%%%%%%%%%%%%
\noindent{\bf Acknowledgements} \quad
%%%%%%%%%%%%%%%%%%%%%%%%%%%%%%%%%%%%%%%%%%%%%%%%%%%%%
The present authors would like to thank
Junichi Wakabayashi for providing them
experimental data of magnetoresistance
measurements 
and for very useful discussion on the
present work.
They also thank Tohru Kawarabayashi
and Mark Lundeberg for helpful comments 
on the present study.
This work is supported in part by
the Grant-in-Aid for Scientific Research (C)
(No.21540397) of Japan Society for
the Promotion of Science.
%%%%%%%%%%%%%%%%%%%%%%%%%%%%%%%%%%%%%%%%%%%%%%%%%%

%%%%%%%%%%% APPENDICES %%%%%%%%%%%%%%%%%%%%%%%%%%%%%%%%
\appendix
%%%%%%%%%%%%%% Appendix A %%%%%%%%%%%%%%%%%%%%%%%%%%%%%
%%%%%%%%%%%%%%%%%%%%%%%%%%%%%%%%%%%%%%%%%%%%%%%%%%%%%%%%
\SSC{Definitions and formulas of special functions}
\label{chap:appendixA}
%%%%%%%%%%%%%%%%%%%%%%%%%%%%%%%%%%%%%%%%%%%%%%%%%%%%%%%%

Here we list the definitions and formulas of special functions
used in the text (see, for instance, \cite{AAR99}).
The gamma function $\Gamma(z)$ is defined by
\begin{equation}
\Gamma(z)=\int_0^{\infty} e^{-u} u^{z-1} du,
\quad \Re z >0.
\label{eqn:Gamma1}
\end{equation}
When $n \in \N_0 \equiv \{0,1,2,\dots\}$,
$\Gamma(n+1)=n!$ (we set $0! \equiv 1$).
The following equality is called 
the duplication formula,
\begin{equation}
\Gamma(2z)=\frac{2^{2z}}{2 \sqrt{\pi}}
\Gamma(z) \Gamma(z+1/2).
\label{eqn:Gamma2}
\end{equation}

The Bessel function $J_{\nu}(z)$ with index $\nu \in \R$
is given by
\begin{equation}
J_{\nu}(z)=\left(\frac{z}{2}\right)^{\nu}
\sum_{n=0}^{\infty} \frac{(-1)^n (z/2)^{2n}}
{n! \Gamma(\nu+n+1)}
\label{eqn:Jnu1}
\end{equation}
for a complex number $z$ which is not negative real.
%When $\nu=0$, it has the following integral representation
%\begin{equation}
%J_0(z)=\frac{2}{\pi} \int_0^1
%\frac{\cos(z u)}{\sqrt{1-u^2}} du.
%\label{eqn:J01}
%\end{equation}
Then the modified Bessel function of the first kind is
defined as
\begin{eqnarray}
I_{\nu}(z) &=& \left\{ \begin{array}{ll}
\displaystyle{ e^{-\nu \pi i/2} J_{\nu}(z e^{\pi i/2})}
\quad &(-\pi < {\rm arg} z \leq \pi/2) \cr
\displaystyle{ e^{3 \nu \pi i/2} J_{\nu}(z e^{-3 \pi i/2})}
\quad &(\pi/2 < {\rm arg} z \leq \pi)
\end{array} \right.
\nonumber\\
&=& \left( \frac{z}{2} \right)^{\nu}
\sum_{n=0}^{\infty} \frac{(z/2)^{2n}}{n ! \Gamma(\nu+n+1)}.
\label{eqn:Inu1}
\end{eqnarray}

For $n \in \N_0$, the Pochhammer symbol is defined as
\begin{eqnarray}
&& (\alpha)_n=\alpha(\alpha+1)(\alpha+2) \cdots (\alpha+n-1), 
\quad n \in \N \equiv \{1,2,3, \dots\},
\nonumber\\
&& (\alpha)_0=1.
\label{eqn:Poch}
\end{eqnarray}
The Gauss hypergeometric function is defined by
\begin{equation}
F(\alpha, \beta, \gamma; z)
=\sum_{n=0}^{\infty} \frac{(\alpha)_n (\beta)_n}{(\gamma)_n}
\frac{z^n}{n!},
\label{eqn:F1}
\end{equation}
and the confluent hypergeometric function is defined as
\begin{eqnarray}
F(\alpha, \gamma;z)
&=& \lim_{\beta \to \infty} 
F(\alpha, \beta, \gamma; z/\beta)
\nonumber\\
&=& \sum_{n=0}^{\infty} 
\frac{(\alpha)_n}{(\gamma)_n} \frac{z^n}{n!}.
\label{eqn:F2}
\end{eqnarray}
The following recurrence relations for $F(\alpha, \beta, \gamma; z)$
are known,
\begin{eqnarray}
\label{eqn:G_rec1}
&& \gamma[F(\alpha, \beta+1, \gamma; z)
- F(\alpha, \beta, \gamma; z)]
=\alpha z F(\alpha+1, \beta+1, \gamma+1; z),
\\
&& \alpha(1-z) F(\alpha+1, \beta, \gamma;z)
+[\gamma-2 \alpha+(\alpha-\beta)z] F(\alpha, \beta, \gamma;z)
\nonumber\\
\label{eqn:G_rec2}
&& \qquad \qquad \qquad \qquad \qquad 
+(\alpha-\gamma) F(\alpha-1, \beta, \gamma;z)=0.
\end{eqnarray}
For $F(\alpha, \beta; z)$,
\begin{eqnarray}
\label{eqn:c_rec1}
&& zF(\alpha+1, \gamma+1;z)
=\gamma[F(\alpha+1, \gamma;z)-F(\alpha, \gamma;z)],
\\
\label{eqn:c_rec2}
&& \alpha F(\alpha+1, \gamma+1; z)
=(\alpha-\gamma) F(\alpha, \gamma+1; z)
+\gamma F(\alpha, \gamma; z)
\end{eqnarray}
are satisfied.
The Gauss hypergeometric function has the
following integral expression,
\begin{equation}
F(\alpha, \beta, \gamma; z)
=\frac{\Gamma(\gamma)}{\Gamma(\beta) \Gamma(\gamma-\beta)}
\int_0^1 du \,
u^{\beta-1} (1-u)^{\gamma-\beta-1}
(1-uz)^{-\alpha}.
\label{eqn:GaussInt}
\end{equation}

The complete elliptic integrals of the first kind, $K(z)$,
and of the second kind, $E(z)$, are defined as
\begin{eqnarray}
K(z) &=& \int_0^{\pi/2} \frac{d \theta}{\sqrt{1-z^2 \sin^2 \theta}}
=\int_0^1 \frac{du}{\sqrt{(1-u^2)(1-z^2 u^2)}},
\nonumber\\
E(z) &=& \int_0^{\pi/2} \sqrt{1-z^2 \sin^2 \theta} d \theta,
\label{eqn:KE1}
\end{eqnarray}
respectively. They are expressed by using the Gauss
hypergeometric functions as
\begin{equation}
K(z)=\frac{\pi}{2} F\left( \frac{1}{2}, \frac{1}{2}, 1; z^2 \right),
\quad
E(z)=\frac{\pi}{2} F\left( -\frac{1}{2}, \frac{1}{2}, 1; z^2 \right).
\label{eqn:KE2}
\end{equation}
The modified Bessel function (\ref{eqn:Inu1}) is expressed
by using the confluent hypergeometric functions as
\begin{equation}
I_{\nu}(z)= \frac{(z/2)^{\nu}}{\Gamma(\nu+1)}
e^{z} F \left( \nu+\frac{1}{2}, 2 \nu+1; -2z \right).
\label{eqn:Inu2}
\end{equation}

%%%%%%%%%%%%% Appendix B %%%%%%%%%%%%%%%%%%%%%%%%%%%%%%%
%%%%%%%%%%%%%%%%%%%%%%%%%%%%%%%%%%%%%%%%%%%%%%%%%%%%%%%%
\SSC{Linear relation between $\rho$ and $\tau^{-1}$}
\label{chap:appendixB}
%%%%%%%%%%%%%%%%%%%%%%%%%%%%%%%%%%%%%%%%%%%%%%%%%%%%%%%%

Let $f_0(k)=[e^{(E(k)-\mu)/(k_{\rm B} T)}+1]^{-1}$
be the equilibrium Fermi distribution function
at temperature $T$ and chemical potential $\mu$.
The relaxation time $\tau(k)$ is 
phenomenologically introduced as
\begin{equation}
\left[ \frac{\partial f(k)}{\partial t} \right]_{\rm S}
=-\frac{\delta f(k)}{\tau(k)}
\quad \mbox{with} \quad \delta f(k)=f(k)-f_0(k).
\label{eqn:tauA1}
\end{equation}
When we consider the long-term limit,
$\partial f(k)/\partial t$-term in the LHS
is zero, and we have
\begin{eqnarray}
\delta f(k) &=& -\frac{e}{\hbar} \tau(k) 
\E \cdot \nabla_k f(k)
\nonumber\\
&\simeq& - \frac{e}{\hbar}
\tau(k) \E \cdot \nabla_k f_0(k).
\label{eqn:tauA2}
\end{eqnarray}
In low temperature limit,
$E=\hbar v_{\rm F} k$ and
$\E \cdot \nabla_k f_0(k)
\simeq \hbar v_{\rm F} |\E| 
\partial f_0(k)/\partial E
\simeq |\E| \delta(k-k_{\rm F})$,
where $k_{\rm F}$ denotes the
Fermi wave-number.
Then in the 2D plane
the current is calculated as follows,
provided $\J$ and $\E$ are parallel;
\begin{eqnarray}
J =|\J| &=&
\int \frac{d^2 k}{(2 \pi)^2}
(-e v_{\rm F}) \delta f(k)
\nonumber\\
&\simeq& \int_0^{\infty} \frac{dk}{2 \pi} k
(-ev_{\rm F}) \left( - \frac{e}{\hbar} \tau(k)
|\E| \delta(k-k_{\rm F}) \right)
\nonumber\\
&=& \frac{e^2 v_{\rm F} k_{\rm F} \tau(k_{\rm F})}
{2 \pi \hbar} |\E|.
\label{eqn:JE}
\end{eqnarray}
Since resistance $\rho$ is defined as
$J=\rho^{-1} |\E|$,
we have the linear relation
between $\rho$ and $\tau^{-1}$ as
$\rho=c \tau^{-1}$
with the coefficient 
$c=2 \pi \hbar/(e^2 v_{\rm F} k_{\rm F})$.
In Eq.(\ref{eqn:rhoB1}) $k_{\rm F}$ is
written as $k$.

%%%%%%%%%%%%%%%%%%%%%%%%%%%%%%%%%%%%%%%%%%%%%%%%%%%%%%%%%%%%
%%%%%%%%% References %%%%%%%%%%%%%%%%%%%%%%%%%%%%%%%%%%%%%%%
%%%%%%%%%%%%%%%%%%%%%%%%%%%%%%%%%%%%%%%%%%%%%%%%%%%%%%%%%%%%

%%%%%%%%%%%%%%%%%%%%%%%%%%%%%%%%%%%%%%%%%%%%%%%%%%%%%%%%%%
\end{document}